\documentclass[twoside,fleqn]{article}
\usepackage{espcrc2,epsf}

\evensidemargin\oddsidemargin
\makeatletter
\long\def\@makefigurecaption#1#2{\vskip 6mm #1. #2\par}
\makeatother
\def\er#1#2{\relax\ifmmode{}^{+#1}_{-#2}\else$^{+#1}_{-#2}$\fi}
\def\erparen#1#2{\relax\ifmmode{}(^{#1}_{#2})\else$(^{#1}_{#2})$\fi}
\def\btorho{\bar B^0{\to}\rho^+ l^- \bar\nu_l}
\def\btopi{\bar B^0{\to}\pi^+ l^- \bar\nu_l}
\def\btokstargamma{\bar B{\to} K^* \gamma}
\def\qsqmax{q^2_{\mathrm{max}}}
\def\vub{|V_{ub}|}
\def\w{\omega}

\def\lsize#1#2{$#1^3\times#2$}
\def\fbstat{f_B^{\mathrm{stat}}}
\def\kcrit{\kappa_{\mathrm{crit}}}
\def\bra#1{\left\langle #1 \right|}
\def\ket#1{\left| #1 \right\rangle}
\def\gev{\,{\rm Ge\kern-0.1em V}}
\def\mev{\,{\rm Me\kern-0.1em V}}
\def\ts{\vrule height3ex depth0pt width0pt}
\def\tts{\vrule height4ex depth0pt width0pt}
\def~{\ifmmode\phantom{0}\else\penalty10000\ \fi}
\newdimen\unit
\def\point#1 #2 #3{\vbox to0pt{\kern-#2\unit
  \hbox{\kern#1\unit$#3$}\vss}
 \nointerlineskip}
\title{Heavy Light Weak Matrix Elements\thanks{Plenary talk presented
at Lattice 96, 14th International Symposium on Lattice Field Theory,
St Louis, USA, June 1996.}\hfill\raisebox{22mm}[0pt][0pt]{\normalsize
SHEP--96--21}}

\author{Jonathan Flynn\address{Department of Physics, University of
Southampton, SO17 1BJ, UK}}
      
\begin{document}

\begin{abstract}
I review the status of lattice calculations of heavy-light weak matrix
elements, concentrating on semileptonic $B$ decays to light mesons,
$\btokstargamma$, the $B$ meson decay constant, $f_B$, and the mixing
parameter $B_B$.
\end{abstract}

\thispagestyle{empty}
\maketitle

\section{INTRODUCTION}

In this review I will focus on calculations of selected matrix
elements involving $b$ quarks. The semileptonic decays $\btopi$ and
$\btorho$ depend on the CKM element $V_{ub}$ which fixes one side of
the unitarity triangle. Lattice calculations open the possibility for
a model independent extraction of this quantity. The rare radiative
decay $\btokstargamma$ is important for the determination of $V_{ts}$
and as a window on new physics. Calculations here are hampered by the
necessity to impose models for form factors. For the $B$ meson decay
constant, the systematic uncertainty arising from continuum
extrapolation is being reduced and quenching effects are being
addressed. Combining with results for the mixing parameter, the
phenomenologically relevant quantity $f_B^2 B_B$ will become better
determined.

\section{SEMILEPTONIC AND RADIATIVE HEAVY-TO-LIGHT DECAYS}

Lattice calculations of form factors are crucial for these decays:
heavy quark symmetry relates different matrix elements and hence form
factors, but the overall normalisation at the zero recoil point $\w =
v{\cdot}v' = 1$ is not fixed by the symmetry as it is for a
heavy-to-heavy transition. Here $v$ and $v'$ are the four-velocities
of the $b$ quark and the light quark it decays into, respectively.

In reviewing results for these decays, I would like to emphasize how
lattice calculations may be used for model independent extractions of
CKM angles.

\begin{figure}
\hbox to\hsize{\hfill\vbox{\offinterlineskip
\epsfxsize=0.7\hsize\epsffile{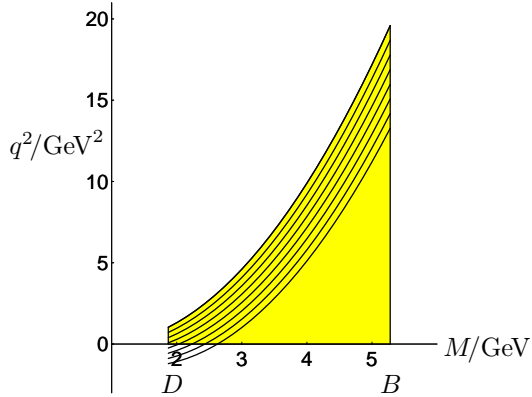}
\unit=0.7\hsize
\point -0.03 0.66 {q^2\!/\!\gev^2}
\point 1 0.17 {M\!/\!\gev}
\point 0.33 0.08 D
\point 0.85 0.08 B
}\hfill}
\caption[]{$q^2$ range for heavy-to-light decays as a function of the
decaying heavy meson mass. Lines of constant $\w$ are shown. The light
final state mass is taken to be $850\mev$, typical of lattice
pseudoscalar or vector meson masses before chiral extrapolation.}
\label{fig:heavy-to-light-scaling}
\end{figure}

The diagram in Fig.~\ref{fig:heavy-to-light-scaling} illustrates the
problem in obtaining form factors for $B$ decays over the full range
of $q^2 = M^2 + m^2 - 2Mm\w$, when a heavy meson of mass $M$ decays to
a light meson of mass $m$, transferring four-momentum $q$ to the
leptons or photon. Calculations using the conventional lattice Dirac
equation (hereafter referred to as the ``conventional'' approach) are
currently performed with heavy quark masses around the charm mass. The
initial heavy meson is given $0$ or $1$ lattice units of
three-momentum, while the light final meson can generally be given up
to two lattice units of spatial momentum, allowing $q^2$ to be varied
from $\qsqmax = (M{-}m)^2$ (where $\w=1$) down to $q^2 <0$ at the $D$
scale. Heavy quark symmetry shows that the the form factors scale
simply with the heavy mass $M$ for fixed $\w$. Table~\ref{tab:hlff}
shows the leading $M$ dependences at fixed $\w$ for the form factors
in the helicity basis~\cite{lms}, where the $J^P$ quantum numbers are
fixed for the $t$-channel. These are multiplied by power series in
$1/M$ to build up the full $M$ dependence.
\begin{table}
\caption[]{Leading $M$ dependence of form factors for heavy-to-light
decays in the helicity basis. The dependence follows from heavy quark
symmetry applied at \emph{fixed} velocity transfer $\w = v\cdot
v'$. Note that only three of the four $A_i$ form factors for $\btorho$
are independent.}
\label{tab:hlff}
\begin{center}
\begin{tabular}{lll}
\hline
\tts\parbox{3em}{Form\\factor$\phantom{\mathrm{g}}$} &
 \parbox{6em}{$t$-channel\\ exchange} &
 \parbox{6em}{Leading $M$\\ dependence} \\[2ex]
\hline
\multicolumn{3}{c}{\ts$B\to\rho l\nu$}\\[0.5ex]
$V$ & $1^-$ & $M^{1/2}$ \\
$A_1$ & $1^+$ & $M^{-1/2}$ \\
$A_2$ & $1^+$ & $M^{1/2}$ \\
$A_3$ & $1^+$ & $M^{3/2}$ \\
$A_0$ & $0^-$ & $M^{1/2}$\\[0.2ex]
\multicolumn{3}{c}{\ts$B\to\pi l\nu$}\\[0.5ex]
$f^+$ & $1^-$ & $M^{1/2}$ \\
$f^0$ & $0^+$ & $M^{-1/2}$ \\[0.2ex]
\multicolumn{3}{c}{\ts$B\to K^*\gamma$}\\[0.5ex]
$T_1$ & $1^-$ & $M^{1/2}$ \\
$T_2$ & $1^+$ & $M^{-1/2}$\\[0.7ex]
\hline
\end{tabular}
\end{center}
\end{table}

As Fig.~\ref{fig:heavy-to-light-scaling} shows, fixed-$\w$ scaling
sweeps all the measured points to a region near $\qsqmax$ for $B$
decays. The problem is then to extrapolate to cover the range back
down to $q^2=0$. This is particularly acute for the radiative decay
$\btokstargamma$ where the two-body final state with a real photon
implies that only the form factors at $q^2=0$ contribute to the
decay. Even when the heavy quark is treated using a static,
non-relativistic or other modified action, the restriction on the
available three-momenta on current lattices ensures that form factor
values are obtained only near $\qsqmax$.

Some assistance is provided by ensuring that any model $q^2$
dependences employed satisfy the requirements of heavy quark symmetry
together with any known constraints relating form factors at
$q^2=0$. For example, $f^+$ and $f^0$ in $B\to\pi$ decays are related
at $q^2=0$, and consistency is achieved by fitting $f^+$ to a dipole
[pole] form and $f^0$ to a pole [constant] form. Heavy quark symmetry
and light flavour $SU(3)$ symmetry relate the form factors for
$\btorho$ and $\btokstargamma$, and models can further relate these
form factors to those for $\btopi$. An overall fit to all the form
factors might then be used. However, it is clearly desirable to avoid
models entirely: I will concentrate on model independent results
below.

Before discussing lattice results, I will summarise the experimental
situation.  The decays $\btorho$ and $\btopi$ are measured using full
reconstruction by the CLEO collaboration, who find~\cite{lkg:ichep96}:
\begin{eqnarray}
B(\btopi)  &=& 1.8(4)(3)(2) \times 10^{-4}, \label{eq:btopi-expt}\\
B(\btorho) &=& 2.5(4)\erparen57(5) \times 10^{-4}.
\end{eqnarray}
The experimental efficiency is model dependent, so the above results
are a combined value using different models. When converted to a
result for $\vub$, the model dependence dominates the errors:
\[
\vub = 3.3(2)\erparen37(7) \times 10^{-3},
\]
to be compared with $\vub = (3.1 \pm 0.8) \times 10^{-3}$ obtained
from the lepton energy spectrum endpoint analysis in inclusive decays,
which is also model dependent.  The possibility of using lattice
results to remove the model dependence from the determination of
$\vub$ is exciting motivation to pursue such studies.

The experimental result for $\btokstargamma$ also comes from CLEO, who
quote~\cite{cleo:ichep96},
\begin{equation}
B(\btokstargamma) = 4.2(8)(6) \times 10^{-5}.
\end{equation}
Combining this with their measurement for the inclusive
decay~\cite{cleo:bsgamma-inclusive},
\begin{equation}
B(b\to s\gamma) = 2.32(57)(35) \times 10^{-4},
\end{equation}
they find~\cite{cleo:ichep96},
\begin{equation}\label{eq:RKstar}
R_{K^*} = {\Gamma(\btokstargamma)\over\Gamma(b\to s\gamma)} =
 0.181(68).
\end{equation}
\begin{table*}
\begin{center}
\caption[]{Lattices used for calculations of heavy-to-light matrix
elements. All simulations are quenched. In the FNAL
case~\cite{fnal:fBfD-lat95}, heavy quarks are treated with the
Fermilab formalism~\cite{simone:lat95}.}
\label{tab:lattices-for-heavy-to-light}
\begin{tabular}{llllllll}
\hline
\ts Label & Ref. & $\beta$ & Lattice size & Cfgs & $a^{-1}/\gev$ &
 Scale set by & Action \\[1ex]
\hline
\ts FNAL & \cite{fnal:fBfD-lat95} & 5.9 &
 \lsize{16}{32} & 300 & 1.78 & 1P--1S in $c\bar c$& SW $c=1.4$ \\
BHSa  & \cite{bhs:bsg} & 6.0 &
 \lsize{16}{39} &  $~39$ & 2.10 & $f_\pi$ & W \\
BHSb  & \cite{bhs:bsg} & 6.0 &
 \lsize{24}{39} &  $~16$ & 2.29 & $f_\pi$ & W \\
LANL  & \cite{lanl:wme-lat95} & 6.0 &
 \lsize{32}{64} & 170 & 2.33 & $m_\rho$ & W \\
APE   & \cite{ape:bsg-clover} & 6.0 &
 \lsize{18}{64} & 170 & 1.96 & $m_\rho$ & SW $c=1$\\
UKQCDa & \cite{ukqcd:hlff} & 6.2 &
 \lsize{24}{48} & $~60$ & 2.73 & $\sqrt\sigma$ & SW $c=1$ \\
UKQCDb & \cite{ukqcd:btorho} & 6.2 &
 \lsize{24}{48} & $~60$ & 2.65 & $m_\rho$ & SW $c=1$ \\
BHSc  & \cite{bhs:bsg} & 6.3 &
 \lsize{24}{61} & $~20$ & 3.01 & $f_\pi$ & W\\
WUP   & \cite{wup:sl-lat95,wup:hl-weak-decays} & 6.3 &
 \lsize{24}{64} & $~60$ & 3.31 & $r_0$ & W \\
ELC   & \cite{elc:hl-semilept} & 6.4 &
 \lsize{24}{60} & $~20$ & 3.7  & $m_\rho $ & W\\[1ex]
\hline
\end{tabular}
\end{center}
\end{table*}
\begin{figure*}
\hbox to\hsize{\hss\vbox{\offinterlineskip
\epsfxsize=0.85\hsize
\epsffile{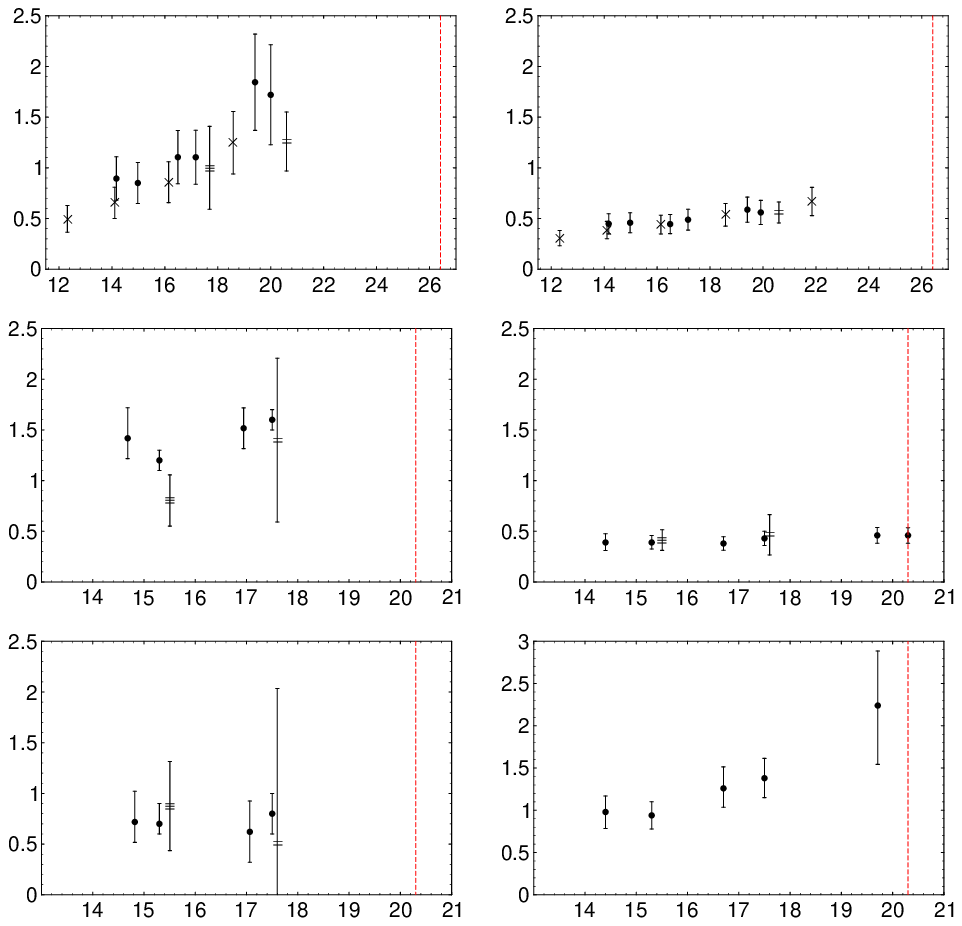}
\unit=0.85\hsize
\point 0.87 0 {q^2/\!\gev^2}
\point 0.055 0.935 {f^+}
\point 0.055 0.61 V
\point 0.055 0.28 {A_2}
\point 0.57 0.935 {f^0}
\point 0.57 0.61 {A_1}
\point 0.57 0.28 {A_0}
}\hss}
\caption[]{Lattice measurements of the form factors for $\btopi$ and
$\btorho$. Points are from ELC~\cite{elc:hl-semilept} ($\equiv$),
APE~\cite{ape:hl-semilept,vl} ($=$), UKQCD~\cite{ukqcd:hlff}
($\bullet$) and FNAL~\cite{simone:lat95,fnal:fBfD-lat95} ($\times$,
preliminary). Systematic errors have been added to the UKQCD and FNAL
points, following~\cite{lpl:btopi-bounds}. The vertical dashed lines
show $\qsqmax$.}
\label{fig:btopirho-form-factors}
\end{figure*}

\subsection{Semileptonic $B \to \pi$}

Lattice results for the $\btopi$ form factors, $f^+$ and $f^0$, are
available from ELC~\cite{elc:hl-semilept}, APE~\cite{ape:hl-semilept,vl},
UKQCD~\cite{ukqcd:hlff} and
Wuppertal-HLRZ~\cite{wup:sl-lat95,wup:hl-weak-decays}, together with
preliminary results from
FNAL~\cite{simone:lat95,fnal:fBfD-lat95}. They are plotted in 
Fig.~\ref{fig:btopirho-form-factors} (the Wuppertal-HLRZ group
extrapolated the form factors at $q^2=0$ only, using various
ans\"atze, so their results are not displayed).

For massless leptons, the decay rate is determined by $f^+$
alone. However, the constraint $f^+(0) = f^0(0)$ (with suitable
conventions) makes lattice measurements of both form factors useful.
One procedure uses dispersive constraints to obtain model independent
information, starting from the two-point correlation function of
$V^\mu = \bar u \gamma^\mu b$,
\begin{eqnarray*}
\Pi^{\mu\nu}(q^2) &=& i \int d^4x\, e^{iq\cdot x} \bra0 \mathrm{T}
 V^\mu(x) V^{\nu\,\dagger}(0) \ket0 \\
 &=& (q^\mu q^\nu\!\!{-}q^2 g^{\mu\nu}) \Pi_T(q^2) + q^\mu q^\nu
\Pi_L(q^2).
\end{eqnarray*}
Because the contributions of intermediate $B^*$, $B\pi$, \ldots states
to the corresponding spectral functions are all positive, the following
inequality holds:
\[
\mathrm{Im}\, \Pi_L(t) \geq \phi(t) |f^0(t)|^2,
\]
where $t=q^2$ and $\phi$ is a known function. A similar inequality
relates $\mathrm{Im}\,\Pi_T$ and $f^+$. Combining the inequalities
with subtracted dispersion relations for $\Pi_{T,L}$ provides bounds
on the form factors in terms of quantities which can be evaluated in
perturbative QCD.

\begin{figure}
\hbox to\hsize{\hss\vbox{\offinterlineskip
\epsfxsize=\hsize\epsffile{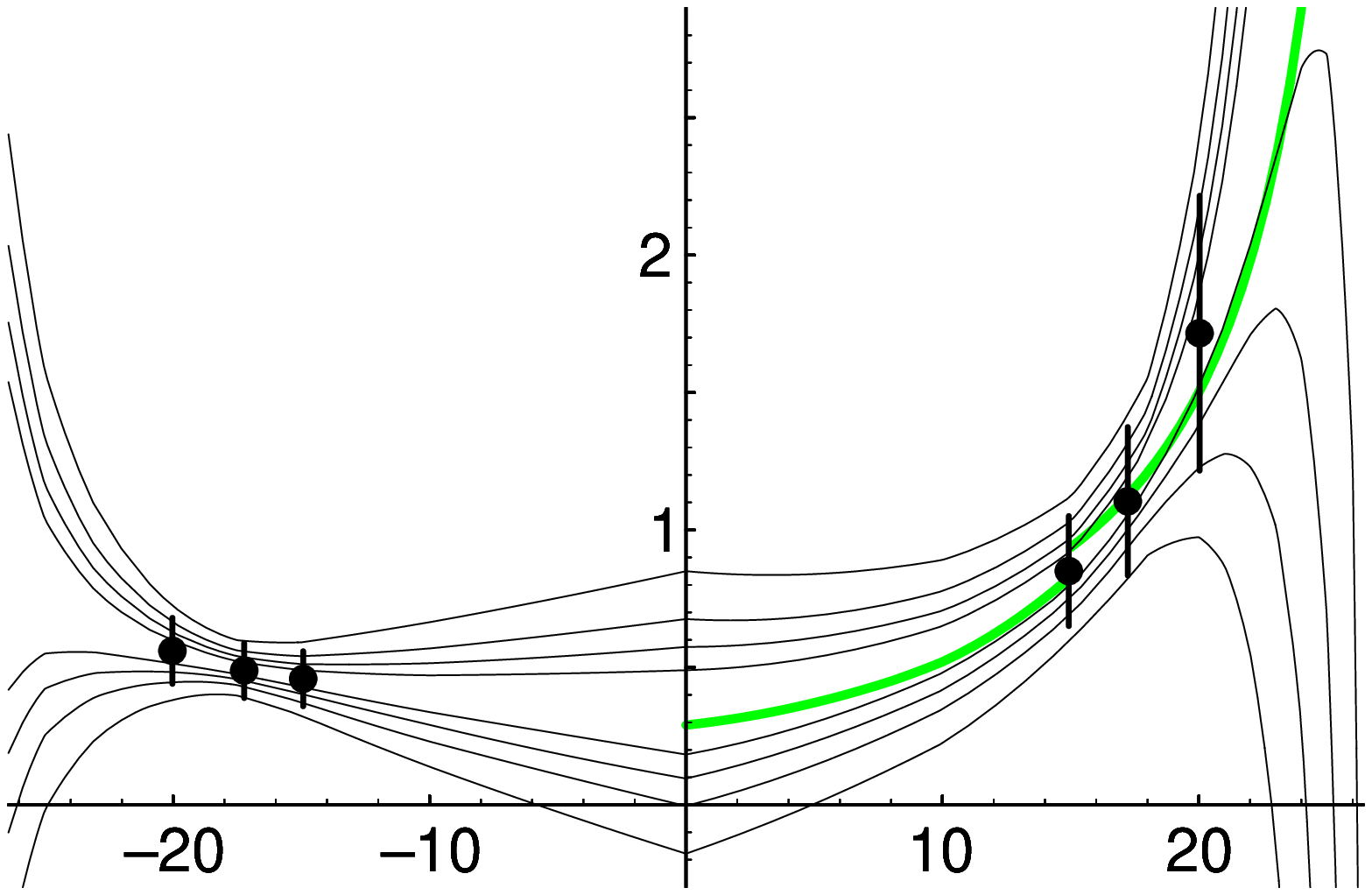}
\unit=\hsize
\point 0.55 0.65 {f^+(q^2)}
\point 0.3 0.65 {f^0(|q^2|)}
\point 0.8 -0.03 {q^2/\gev^2}
}\hss}
\caption[]{Bounds on the form factors $f^+$ and $f^0$ for $\btopi$
obtained from dispersive constraints~\cite{lpl:btopi-bounds}. The
points displayed are from UKQCD~\cite{ukqcd:hlff}, with added
systematic errors, and the shaded band is the prediction of a
light-cone sum rule calculation~\cite{bbkr:B-Bstar-pi-couplings}.}
\label{fig:btopi-bounds}
\end{figure}

Known values of the form factors can be incorporated to tighten the
bounds. Lellouch~\cite{lpl:btopi-bounds} has shown how to incorporate
the $f^+(0)=f^0(0)$ constraint together with imperfectly known values
of the form factors, typical of lattice results with errors, to obtain
families of bounds with varying confidence levels. A set of such
bounds are shown in Fig.~\ref{fig:btopi-bounds} together with the
UKQCD~\cite{ukqcd:hlff} lattice measurements used to obtain them. In
the figure, $f^0$ and $f^+$ are plotted back-to-back, showing the
effect of imposing the constraint at $q^2=0$.

In Table~\ref{tab:btopi-results} the bounds have been used to give
ranges of values for the decay rate $\Gamma(\btopi)$ in units of
$\vub^2\,\mathrm{ps}^{-1}$ together with values for the form factor
$f^+$ at $q^2=0$. When combined with the experimental result for the
decay rate in Eq.~(\ref{eq:btopi-expt}), one can extract $\vub$
with about 35\% theoretical error. Although this result is not very
precise, the procedure used relies only on lattice calculations of
matrix elements and heavy quark symmetry, together with perturbative
QCD and analyticity properties in applying the dispersive
constraints. There is no model dependence. Development and application
of improved lattice actions for heavy quarks may in future remove the
need for heavy quark symmetry in the extrapolation from $D$ to $B$
mesons. It is tempting to consider that lattice results could
eventually supplant models in the experimental efficiency
determinations, thereby removing model dependence from the
experimental results.

Also shown in Table~\ref{tab:btopi-results}, for comparison, are
values obtained from lattice calculations where assumed $q^2$
dependences have been imposed. For ELC~\cite{elc:hl-semilept} and
APE~\cite{ape:hl-semilept}, one value of $f^+$ has been used, at the
given value of $q^2$, fitted to a single pole form with pole mass
$m_\mathrm{p}$. The UKQCD result~\cite{ukqcd:hlff} is obtained from a
combined dipole/pole fit to all measured $f^+$/$f^0$ points. Note that
the UKQCD points have statistical errors only and have not been
chirally extrapolated---they correspond to a pion mass of around
$800\mev$ (a similar caveat applies for the
FNAL~\cite{simone:lat95,fnal:fBfD-lat95} results for $f^+$ and
$f^0$). The results given~\cite{drb} have used these values as though
they applied to the physical pion. In obtaining bounds based on these
points, Lellouch~\cite{lpl:btopi-bounds} added a conservatively
estimated systematic error including terms to account for the chiral
extrapolation (this error has been added to the UKQCD and FNAL points
plotted in Fig.~\ref{fig:btopirho-form-factors}). Needless to say,
improved lattice results can be used as input for the bounds once they
become available.
\begin{table}
\caption[]{Results for $\btopi$ from dispersive constraints applied to
lattice results~\cite{lpl:btopi-bounds}, together with results
obtained using ans\"atze for the form factor $f^+$. Collaboration
labels in the left hand column refer to lattices described in
Table~\ref{tab:lattices-for-heavy-to-light}. The decay rates are
values for the quantity $\Gamma(\btopi)/\vub^2 \mathrm{ps}^{-1}$}
\label{tab:btopi-results}
\begin{center}
\begin{tabular}{llll} \hline
\ts  & Rate & $f^+(0)$ & \\[1ex]
\hline
\ts Dispersive  & $2.4$--$28$ & $-0.26$--$0.92$ & 95\% CL\\
Constraint & $3.6$--$17$ & $0.00$--$0.68$ & 70\% CL\\
\cite{lpl:btopi-bounds} & $4.8$--$10$ & $0.18$--$0.49$ & 30\% CL\\[1ex]
\hline
\ts ELC & $9\pm 6$ & 0.10--0.49 & \\
\multicolumn{4}{l}{$q^2{\simeq} 18\gev^2$, pole fit,  
$m_\mathrm{p}{=}5.29(1)\gev$ \cite{elc:hl-semilept}}\\[1ex]
\hline
\ts APE & $8\pm 4$ & 0.23--0.43 & \\
\multicolumn{4}{l}{$q^2{\simeq} 20.4\gev^2$, pole fit,
$m_\mathrm{p}{=}5.32(1)\gev$ \cite{ape:hl-semilept}}\\[1ex]
\hline
\ts UKQCDa & $7 \pm 1$ & 0.21--0.27 & \\
\multicolumn{4}{l}{dipole/pole fit to $f^+$/$f^0$
 \cite{ukqcd:hlff,drb}}\\[1ex]
\hline
\end{tabular}
\end{center}
\end{table}

Discretisation errors are an important source of systematic
uncertainty~\cite{simone:lat95}. Mass dependent errors affect the
crucial extrapolation from $D$ to $B$ mesons in calculations using
standard Wilson or Sheikholeslami-Wohlert (SW) quarks. Momentum
dependent discretisation errors and increasing statistical uncertainty
also limit the range of spatial momenta and hence $q^2$ that can be
probed.

Discretisation errors in the heavy-mass scaling were studied by the
FNAL group~\cite{simone:lat95,fnal:fBfD-lat95}, using the Fermilab
improvement formalism. They studied $\bra\pi V^\mu \ket P/ \sqrt{2M
E_\pi}$, where $P$ is a pseudoscalar meson of mass $M$. This quantity
should be constant in the $M\to\infty$ limit. The matrix elements were
computed for $D$ and $B$ mesons as well as in the static limit. The
improvement prescription resulted in little dependence on $1/M$ for a
range of pion momenta.

Momentum-dependent discretisation errors have been studied by a
FNAL-Illinois-Hiroshima group~\cite{simone:lat96} also using the
Fermilab formalism for heavy quarks. From a study of the axial current
matrix element between a pseudoscalar and the vacuum, they estimate
that momentum dependent errors are less than 20\% for light mesons
with $|\mathbf{p}| < 1.2\gev$ on a relatively coarse $\beta=5.7$
lattice.

Improvement will clearly help in the determination of reliable
phenomenological results from lattice calculations. It will be
particularly interesting to test the benefit of the $O(a)$ improvement
program outlined by the ALPHA collaboration~\cite{alpha:lat96}.

I note finally in this section that chiral extrapolations are severe
for the $B\to\pi$ matrix elements. As the pion mass approaches its
physical value, the $B^*$-pole and the beginning of the $B\pi$
continuum move very near the upper endpoint of $q^2$. The form factors
may vary rapidly with the pion mass near $\qsqmax$. For this reason,
the lattice calculations of semileptonic $B\to\rho$ decay are
currently most reliable and I now turn to this.

\subsection{Semileptonic $B \to \rho$}

To avoid models for the $q^2$ dependence of the form factors, we use
the lattice results directly. The lattice can give the
differential decay rate $d\Gamma/dq^2$, or the partially integrated
rate, in a $q^2$ region near $\qsqmax$ up to the unknown factor
$\vub^2$. For example, UKQCD~\cite{ukqcd:btorho} parametrised the
differential decay rate near $\qsqmax$ by,
\begin{equation}
{d\Gamma\over dq^2} = \mathrm{const} |V_{ub}|^2 q^2 \lambda^{1/2} a^2
\big(1+b[q^2{-}\qsqmax]\big),\label{eq:dGdqsq}
\end{equation}
where $\lambda$ is the usual phase space factor and $a$ and $b$ are
constants. The constant $a$ plays the role of the Isgur-Wise function
evaluated at $\w=1$ for heavy-to-heavy transitions, but in this case
there is no symmetry to determine its value at leading order in the
heavy quark effective theory.

For massless leptons, the differential decay rate depends on the $V$,
$A_1$ and $A_2$ form factors, but $A_1$ is the dominant contribution
near $\qsqmax$ and is also the best measured on the lattice, as shown
in Fig.~\ref{fig:btopirho-form-factors}.

\begin{figure}
\hbox to\hsize{\hfill\epsfxsize=0.9\hsize
\epsffile[25 37 516 505]{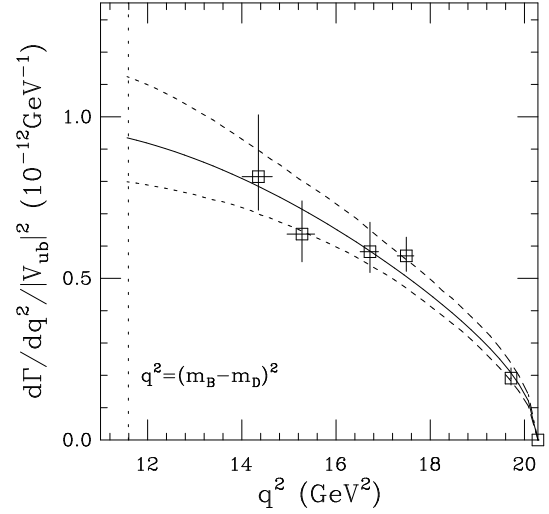}
\hfill}
\caption[]{UKQCD calculation of the differential decay rate for
$\btorho$ as a function of $q^2$~\cite{ukqcd:btorho}.}
\label{fig:ukqcd-btorho}
\end{figure}

Fig.~\ref{fig:ukqcd-btorho} shows UKQCD results for the differential
decay rate together with a fit to the parametrisation in
Eq.~(\ref{eq:dGdqsq}). Because of phase space suppression, the
point at $\qsqmax$ does not influence the fit, and so gives another
determination of $a^2$. The values found are~\cite{ukqcd:btorho},
\[
a^2 = \cases{21(3) \gev^2&fit to 5 points,\cr
23(2) \gev^2&$\qsqmax$ point,\cr}
\]
leading to:
\[
a = 4.6 \er{0.3}{0.4} \mathrm{(stat)} \pm 0.6 \mathrm{(syst)} \gev.
\]
Discounting experimental errors, this result will allow determination
of $\vub$ with a theoretical uncertainty of 10\% statistical and 12\%
systematic.
With full reconstruction of events, CLEO are beginning to extract the
differential decay distributions~\cite{lkg:ichep96}. The Babar
experiment, with unequal beam energies allowing verification
that the $\rho$ and lepton originate from the same vertex, should do
even better.

The UKQCD results for $V$, $A_1$ and $A_2$ agree very well with a
light cone sum rule (LCSR) calculation of Ball and
Braun~\cite{ball:lcsr-moriond}. The agreement is perhaps
fortuitous. More interestingly, LCSR calculations predict that all the
form factors for heavy-to-light decays have the following heavy mass
dependence at $q^2=0$~\cite{bbkr:B-Bstar-pi-couplings,cz:lcsr-orig}:
\begin{equation}
f(0) = {1\over M^{3/2}} (a_0 + a_1/M + a_2/M^2 + \cdots).
\label{eq:lcsr-ff-zero}
\end{equation}
The leading $M$ dependence comprises $\sqrt{M}$ from the heavy state
normalisation together with the behaviour of the leading twist light
cone wavefunction. For $A_1(0)$ the LCSR result
is~\cite{ball:lcsr-moriond} $A_1(0) \simeq 0.26$. UKQCD fitted the
heavy mass dependence of $A_1(0)$ to the form in
Eq.~\ref{eq:lcsr-ff-zero} and found~\cite{ukqcd:btorho}
\[
A_1(0) = \cases{0.18\pm0.02&$a_0$ and $a_1 \neq 0$\cr
0.22\er{0.04}{0.03}&$a_0$, $a_1$ and $a_2 \neq 0$\cr}
\]
Pole fits for $A_1$ (and dipole/pole for $f^+$/$f^0$ for $B\to\pi$
decays) have leading $M^{-3/2}$ behaviour at $q^2=0$, so we can also
compare with other lattice results:
\[
A_1(0) = \cases{0.22\pm0.05&ELC \cite{elc:hl-semilept}\cr
0.24\pm0.12&APE \cite{ape:hl-semilept}\cr
0.27\er74&UKQCD \cite{ukqcd:btorho}\cr}
\]
The Wuppertal-HLRZ group~\cite{wup:sl-lat95,wup:hl-weak-decays} tried
various forms for the heavy mass dependence of $A_1(0)$, although none
had a leading $M^{-3/2}$ dependence.

\subsection{Rare radiative $B \to K^* \gamma$}

This decay was discussed in some detail by A.~Soni at Lattice
95~\cite{soni:lat95} so my comments will be brief. In
Table~\ref{tab:btokstargamma} I summarise the available lattice
results, all from quenched simulations for the matrix element
\begin{equation}\label{eq:bkstarME}
\langle K^*(k,\eta) | \overline{s} \sigma_{\mu\nu} q^\nu b_R
 | B(p) \rangle
\end{equation}
which is parameterised by three form factors, $T_i$, $i=1,2,3$. For
the decay rate the related values $T_1(0)$ and $T_2(0)$ are
needed. Suitably defined, they are equal, so in the table I quote a
single value $T(0)$, together with the directly measured
$T_2(\qsqmax)$. The results are classified according to the leading
$M$ dependence of the form factor at $q^2=0$ which is governed by the
model used to fit the $q^2$ dependence. Dipole/pole forms for
$T_1$/$T_2$ give $M^{-3/2}$ behaviour and pole/constant forms give
$M^{-1/2}$. The table shows that the results agree when the same
assumptions are made. All groups find that $T_2$ has much less $q^2$
dependence than $T_1$, but the overall forms cannot be decided, so a
phenomenological prediction is elusive.
\begin{table}
\caption[]{Lattice results for $\btokstargamma$. The labels in the
left hand column refer to the lattices described in
Table~\ref{tab:lattices-for-heavy-to-light}.}
\label{tab:btokstargamma}
\begin{center}
\begin{tabular}{llll}\hline
\ts & \multicolumn{2}{c}{$T(0)$} \\[0.5ex]
\cline{2-3}
\ts & $M^{-3/2}$ & $M^{-1/2}$ & $T_2(\qsqmax)$
\\[0.5ex]
\hline
\ts BHSb & 0.10(3) &         & 0.33(7) \\
LANL   & 0.09(1) & 0.24(1) & \\
APE    & 0.09(1)(1) & 0.23(2)(2) & 0.23(2)(2) \\
UKQCDa & 0.15\erparen76 & 0.26\erparen21 & 0.27\erparen21 \\
BHSc   & & & 0.30(3)\\[0.5ex]
\hline
\end{tabular}
\end{center}
\end{table}

Additional long distance contributions may not be negligible so the
matrix element of Eq.~(\ref{eq:bkstarME}) may not give the true
decay rate~\cite{gp:longdist1,gp:longdist2,abs}. Once the $q^2$
dependence of the form factors is known, lattice calculations of the
ratio $R_{K^*} = \Gamma(\btokstargamma)/\Gamma(b\to s\gamma)$ can be
compared to the experimental result in Eq.~(\ref{eq:RKstar}) to
test for long distance effects.

Heavy quark symmetry, combined with light flavour $SU(3)$ symmetry,
relates the form factors for $\btorho$ and
$\btokstargamma$~\cite{iw:hqet,gmm} in the infinite heavy quark mass
limit. On the lattice these relations can be tested using identical
light quarks. UKQCD~\cite{ukqcd:btorho} showed that the ratios
$V/2T_1$ and $A_1/2T_2$ both satisfied the heavy quark symmetry
prediction of unity in the infinite mass limit. A combined fit of the
pseudoscalar to vector form factors consistent with heavy quark
symmetry could help resolve the ambiguity in the $q^2$ dependence of
the $\btokstargamma$ form factors.

\section{LEPTONIC DECAY CONSTANTS OF HEAVY-LIGHT PSEUDOSCALARS}

Leptonic decay constants of heavy light systems were fully reviewed by
C.~Allton at Lattice 95~\cite{allton:lat95}, so here I report new
results. To establish notation, the decay constant is determined by
lattice calculation of the dimensionless quantity $Z_L$ according to,
\[
f_P \sqrt{M_P/2} = Z^{\mathrm{ren}}Z_L a^{-3/2}
\]
where $Z^{\mathrm{ren}}$ is the renormalisation constant required to
match to the continuum and $a$ is the lattice spacing.

\subsection{Conventional Methods}

New results were shown by the MILC~\cite{milc:fb-lat96} and
JLQCD~\cite{jlqcd:fb-lat96} collaborations. Both groups perform
continuum extrapolations from results at several $\beta$'s.

\begin{figure}
\hbox to\hsize{\hss\epsfxsize=0.85\hsize
\epsffile{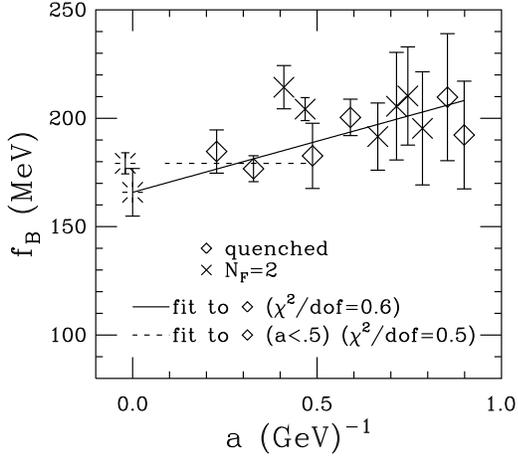}\hss}
\caption{Continuum extrapolation of $f_B$ by
MILC~\cite{milc:fb-lat96}. The dashed line is a constant fit for
quenched results with $\beta\geq6.0$.}
\label{fig:milc-fb}
\end{figure}

MILC have six sets of quenched configurations for $\beta$ in the range
$5.7$--$6.52$, together with six sets of two-flavour dynamical
staggered fermion configurations with $5.445 \leq \beta \leq 5.7$. A
Wilson action is used for the valence quarks, with a hopping parameter
expansion~\cite{ukqcd:hopping} for the heavy quark for easy simulation
of a range of masses.  The heavy quark has a Kronfeld-Lepage-Mackenzie
normalisation, $\sqrt{1 - 6\tilde\kappa}$, together with a shift from
the pole mass to the kinetic mass for the heavy meson.  The continuum
extrapolation should deal with remaining $O(a)$ errors.
Fig.~\ref{fig:milc-fb} shows this extrapolation for $f_B$. MILC's
(preliminary) results are:
\begin{eqnarray*}
f_B &=& 166(11)(28)(14)\mev \\
f_{B_s} &=& 181(10)(36)(18)\mev \\
f_{B_s}/f_B &=& 1.10(2)(5)(7)\\
f_D &=& 196(9)(14)(8)\mev \\
f_{D_s} &=& 211(7)(25)(11)\mev \\
f_{D_s}/f_D &=& 1.09(2)(5)(5)
\end{eqnarray*}
The central values come from the quenched results using a linear
extrapolation in $a$ with the scale set by $f_\pi$ and linear chiral
extrapolations (in $1/\kappa$). The heavy mass dependence is
determined by a fit for heavy meson masses in the range $1.5\gev < M <
4\gev$ combined with a point in the static limit. The first error
quoted is a combination of the statistical error and that coming from
choice of fitting procedure, the second error is remaining systematic
errors within the quenched approximation and the final error is for
quenching.

Improved covariant fitting methods have changed the results compared
to their values at Lattice 95~\cite{milc:fB-lat95}, although within
errors. In particular the difference between quenched and unquenched
results is less dramatic, but still looks significant (especially for
$f_{B_s}$), and indicates an increase in the value of the decay
constant for full QCD.

JLQCD~\cite{jlqcd:fb-lat96} have results from three $\beta$ values,
$5.9$, $6.1$ and $6.3$ with quenched configurations using the Wilson
action for both light and heavy quarks. They study different
prescriptions for reducing the $O(ma)$ scaling violations associated
with the heavy quark and aim to show that results from all
prescriptions converge in the continuum limit.  Smeared wavefunctions
are determined for each available combination of heavy and light quark
kappa values. Results are quoted using the charmonium 1S--1P splitting
to set the scale.

\begin{figure}
\hbox to\hsize{\hss\epsfxsize=\hsize
\epsffile{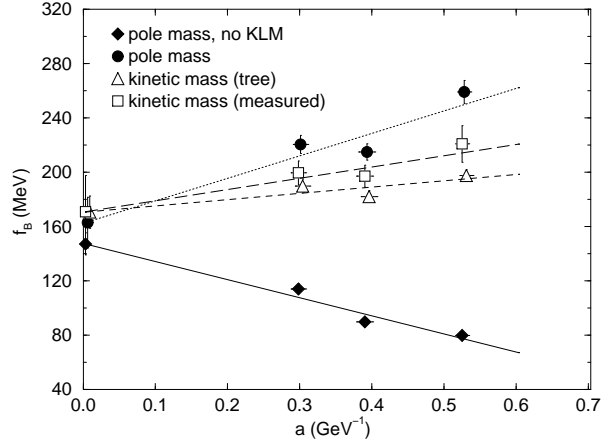}\hss}
\caption{Continuum extrapolation of $f_B$ by
JLQCD~\cite{jlqcd:fb-lat96} using different prescriptions.}
\label{fig:jlqcd-fb}
\end{figure}

JLQCD apply four different procedures to extract the decay
constant. The first uses the traditional $\sqrt{2\kappa}$ quark field
normalisation and the meson pole mass. Three further methods use a
Kronfeld-Lepage-Mackenzie normalisation combined with, (a) setting the
meson mass from the pole mass, (b) pole mass shifted to kinetic mass
at tree level, or (c) measured kinetic mass. For all four methods, a
linear extrapolation in $a$ gives the continuum result. This is shown
for $f_B$ in Fig.~\ref{fig:jlqcd-fb}.  The preliminary results
are~\cite{jlqcd:fb-lat96}:
\begin{eqnarray*}
f_B &=& 163(19)\erparen{~8}{16}\mev \\
f_{B_s} &=& 174(12)\erparen{17}{19}\mev \\
f_D &=& 191(8)\erparen48\mev \\
f_{D_s} &=& 210(6)\erparen{~6}{12}\mev
\end{eqnarray*}
where the first error is statistical and the second is from the spread
over the four methods above.

The MILC and JLQCD results are included in
Table~\ref{tab:pscalar-decay-consts} which summarises determinations
of heavy-light pseudoscalar meson decay constants and their
ratios. MILC and JLQCD are (currently) in pleasing agreement. From the
table, and allowing for an increase in the value from unquenching, I
quote a global result for $f_B$:
\[
f_B = 175 \pm 25 \mev.
\]
\begin{table*}
\caption[]{Summary of results for values and ratios of leptonic decay
constants of pseudoscalar mesons. Statistical and systematic errors
have been combined in quadrature. The results from
MILC~\cite{milc:fb-lat96} include a systematic error for
quenching. Results from APE~\cite{ape:fBfD-lat93} and
FNAL~\cite{fnal:fBfD-lat95} show statistical errors
only. FNAL~\cite{fnal:fBfD-lat95} results are preliminary and use a
nominal value for the strange quark mass to determine $f_{B_s}$ and
$f_{D_s}$. Results with no quoted values for $\beta$ or $a^{-1}$ come
from more than one lattice: PSI-WUP~\cite{psiwup:zpc94},
MILC~\cite{milc:fb-lat96} and JLQCD~\cite{jlqcd:fb-lat96} perform a
continuum extrapolation using their own data.}
\label{tab:pscalar-decay-consts}
\begin{center}
\begin{tabular}{llllllllll}
\hline
\tts & Ref & $\beta$ & $\displaystyle {a^{-1}\over\gev}$ &
$\displaystyle {f_B\over\mev}$ & $\displaystyle {f_{B_s}\over\mev}$ & 
$\displaystyle {f_{B_s}\over f_B}$ &
$\displaystyle {f_D\over\mev}$ & $\displaystyle {f_{D_s}\over\mev}$ & 
$\displaystyle {f_{D_s}\over f_D}$ \\[2ex] \hline
\ts FNAL & \cite{fnal:fBfD-lat95} & 5.9 & & 188\erparen64 &
207\erparen32 & & 220\erparen45 & 239\erparen34 \\
DeG-L & \cite{degrand-loft} & 6.0 & 1.9 & & & & 190(33) & 222(16)
& 1.17(22) \\
APE & \cite{ape:fBfD-lat93} & 6.0 & & 197(18) & & & 218(9) & 240(9) &
1.11(1) \\
UKQCD & \cite{ukqcd:fBfD} & 6.0 & 2.0 &
176\erparen{41}{28} & & 1.17(12) &
199\erparen{30}{24} & & 1.13\erparen67 \\
LANL & \cite{lanl:decay-consts} & 6.0 & 2.33 & & & &
229\erparen{21}{17} & 260\erparen{27}{22} &
1.14(2) \\
BDHS & \cite{bdhs:88} & 6.1 & 2 & & & & 174(53) & 234(72) &
1.35(22) \\
UKQCD & \cite{ukqcd:fBfD} & 6.2 & 2.7 & 160\erparen{53}{20} &
194\erparen{62}{10} & 1.22\erparen43 &
185\erparen{42}{~8} & 212\erparen{46}{~8} &1.18(2) \\
BLS & \cite{bls} & 6.3 & 3.0 & 187(38) & 207(41) & 1.11(6) &
208(38) & 230(36) & 1.11(6) \\
ELC & \cite{elc:npb92} & 6.4 & 3.3 & & & & 210(40) & 230(50) \\[1.5ex]
ELC & \cite{elc:88} & & & & & & 194(15) \\
ELC & \cite{elc:npb92} & & & 205(40) & & 1.08(6) \\
PSI-WUP & \cite{psiwup:fBfD} & & & & & & 198(17) & 209(18) \\
LANL & \cite{lanl:decay-consts} & & & & & & 186(29) & 218(15) \\[1.5ex]
PSI-WUP & \cite{psiwup:zpc94} & & $a{\to}0$ & 180(50) & & 1.09(5) &
170(30) & & 1.09(5) \\
MILC & \cite{milc:fb-lat96} & & $a{\to}0$ & 166(33) & 181(41)
& 1.10(9) & 196(18) & 211(28) &
1.09(7) \\
JLQCD & \cite{jlqcd:fb-lat96} & & $a{\to}0$ & 163\erparen{21}{25} &
174\erparen{21}{22} & & 191\erparen{~9}{11} & 210\erparen{~8}{13}\\[0.5ex]
\hline
\end{tabular}
\end{center}
\end{table*}

\subsection{Decay Constants from NRQCD}

The SGO collaboration have calculated matrix elements for determining
pseudoscalar and vector $B$ meson decay
constants~\cite{sgo:fB-dynam-W,sgo:fB-dynam-SW-TI} using both Wilson
and tadpole-improved SW light quarks combined with
NRQCD heavy quarks. They have $100$ unquenched configurations with
$n_f=2$ flavours of dynamical staggered quark ($m_{\mathrm{sea}}a =
0.01$) at $\beta = 5.6$. The NRQCD action is corrected to $O(1/M)$ and
matrix elements are calculated for extra operators appearing at
$O(1/M)$ in the matching of NRQCD currents onto full QCD vector and
axial vector currents\footnote{See
Refs.~\cite{sgo:fB-dynam-W,sgo:fB-dynam-SW-TI} for details and
qualifications.}.

Appropriate renormalisation constants are not yet available, however,
for the matching to full QCD, so I do not quote here any values for
decay constants.  Accepting this limitation, the slope of the
pseudoscalar decay constant with respect to $1/M$ is calculated, that
is, the term $c_P$ in the expansion
\[
f\sqrt{M} = \mathrm{const}(1 + c_P/M + \cdots).
\]
The results are,
\[
c_P = \cases{-1.35(15) a^{-1} = -2.8(5)\gev&Wilson,\cr
              -1.0(2) a^{-1} \sim -2\gev&TI SW,\cr}
\]
to be compared to slopes of around $1\gev$ found using conventional
methods. Study of a spin average of pseudoscalar and (appropriately
defined) vector meson decay constants shows that just over 90\% of the
effect comes from the heavy quark kinetic energy term. Analysing the
effect of the kinetic term on wavefunctions and energies in the static
limit (from a spinless relativistic quark model) to first order shows
that a slope of $1$--$2 \gev$ is natural, and also predicts that the 1S
state energy should rise slowly with $1/M$, with the 2S energy rising
faster~\cite{cthd:lat96}.

It will be very interesting to see the slope and decay constant
values once the correct renormalisation constants are included.

\subsection{$\fbstat$ from Bermions}

The APETOV group have results for $\fbstat$ obtained using a
pseudofermion method~\cite{bermions2}. In standard static quark
methods, smeared sources must be tuned to optimise the projection onto
the ground state and avoid noise.  Moreover, the numerical expense of
inverting the Dirac operator restricts the set of available light
quark propagators to have one endpoint fixed. In contrast, APETOV
determine the light propagator by Monte Carlo inversion, allowing them
to average the static correlator over \emph{all} points without
requiring smeared sources. The additional average over lattice points
helps overcome the extra noise inherent in the Monte Carlo
inversion.  Further averaging over the gauge links is also employed.

The bermion action is $\sum_x |Q \phi(x)|^2$, where $x$ is a lattice
point, $U$ is the gauge field, $\phi$ the scalar bermion field and $Q$
is given by $Q\phi(x) = \gamma_5 D\phi(x)$, where $D$ is the usual
Wilson lattice Dirac operator. The bermions are thermalised in each
gauge configuration and the pseudofermion two-point function then
gives $(Q^2)^{-1}$.  Remultiplying by $Q$ and $\gamma_5$ allows the
usual light quark Dirac propagator to be determined between any
lattice points. The correlator of two local heavy-light bilinears with
a given time separation is then averaged over all points to determine
$Z_L$.  In practice, a further correlator using a $(Q^2)^{-1}$ bermion
propagator with the static propagator, which better isolates the
lowest lying state, needs to be analysed concurrently.

The method has been applied to two \lsize{16}{32} lattices with
30 gauge configurations each at $\beta = 5.7$ and $6.0$, with the
results~\cite{dediv}:
\[
Z_L = \cases{\cases{0.477(31)&$\kcrit$\cr
             0.582(11)&$\kappa_\mathrm{strange}$\cr}&$\beta=5.7$\cr
             \cases{0.188(37)&$\kcrit$\cr
             0.216(13)&$\kappa_\mathrm{strange}$\cr}&$\beta=6.0$\cr}
\]
The $\beta=6.0$ chirally extrapolated result agrees well with world
results for $Z_L$. At $\beta =5.7$ the result is below the
FNAL~\cite{fnal:staticB} value $Z_L = 0.564(28)$, but is on a larger
lattice and has comparable errors with one third the number of gauge
configurations.

APETOV have extended the method by making the bermions
dynamical~\cite{bermions3}. Each bermion flavour corresponds to $-2$
fermion flavours. $Z_L$ is calculated for different $n_b=-2n_f$,
tuning $\beta$ to keep the ratio $R = m_\pi^2/m_\rho^2$ constant,
starting with $\beta=5.7$ for $n_b=0$. Including a one-loop
perturbative value for $Z^{\mathrm{ren}}$, the result for the ratio of
$Z_L Z^{\mathrm{ren}}$ for 3 flavours to 0 flavors is $1.14(2)$ for
$R=0.6$ and $1.16(4)$ for $R=0.5$, with the effect increasing to about
20\% in the chiral limit. This offers more evidence for the increase
of $f_B$ once dynamical effects are included.

\section{Results for $B$ Mixing}

\begin{table*}
\caption[]{Lattice results for the $B_d$ meson mixing parameter
$B_B$. Lattice parameters are quoted except where the authors have
performed a continuum extrapolation from several results, indicated by
$a{\to}0$ in the $a^{-1}$ column. The column labelled ``Action'' shows
the action used for the light and heavy quarks respectively. The $b$
quark mass $m_b$ has been set to $5\gev$: where a result has been
quoted at a different scale, the scale $\mu$ and $B_B(\mu)$ have
been listed. The results headed ``Static fitted'' are obtained by
extrapolating $1/M \to 0$ where $M$ is the heavy-light meson
mass. Results in oblique type have been obtained from the authors'
values using one-loop scaling with $5$ flavours and
$\Lambda_{(5)\overline{\mathrm{MS}}}=130\mev$.}
\label{tab:B_B}
\begin{center}
\begin{tabular}{lllllllllll}\hline
\tts & Ref & $\beta$ & & Cfgs & $\displaystyle {a^{-1}\over\gev}$ &
Action & $\displaystyle {\mu\over\gev}$ &
$B_B(\mu)$ & $B_B(m_b)$ & $\hat B_B$ \\[2ex] \hline
\multicolumn{11}{l}{\ts Static} \\[0.5ex]
KEN & \cite{ken:lat96} & 6.0 & \lsize{20}{30} & $~32$ & 2.1 & W--stat
& 4.33 & 0.98(5) & \sl 0.97(5) & \sl 1.45(8) \\
APE & \cite{ape:lat96} & 6.0 & \lsize{24}{40} & 600 & 2.0 & SW--stat 
& & & \sl 0.82(4) & 1.21(6) \\
UKQCD & \cite{ukqcd:staticB} & 6.2 & \lsize{24}{48} & $~60$ & 2.9 & 
SW--stat & & & 0.69(4) & 1.02(6) \\[1ex]
\hline
\multicolumn{11}{l}{\ts Static fitted}\\[0.5ex]
ELC & \cite{elc:npb92} & 6.4 & \lsize{24}{60} & $~20$ & 3.7 & W--W & 3.7 &
0.90(5) & \sl 0.88(5) & \sl 1.30(7) \\
BS & \cite{soni:lat95} & \rlap{5.7--6.3} & & & $a{\to}0$ & W--W & 2 
& 1.04(5) & \sl 0.96(5) & \sl 1.42(7) \\[1ex]
\hline
\multicolumn{11}{l}{\ts Conventional}\\[0.5ex]
BBS & \cite{bbs:lat96,tblum} & 5.7 & \lsize{16}{33} & 100 & 1.45 &
W--W & 2 & 0.96(5) & \sl 0.89(5) & \sl 1.32(7) \\
BBS & \cite{bbs:lat96,tblum} & 6.0 & \lsize{16}{39} & $~60$ & 2.06 &
W--W & 2 & 0.96(5) & \sl 0.89(5) & \sl 1.32(7) \\
BBS & \cite{bbs:lat96,tblum} & 6.0 & \lsize{24}{39} & $~40$ & 2.22 &
W--W & 2 & 0.98(4) & \sl 0.90(3) & \sl 1.34(5) \\
JLQCD & \cite{jlqcd:lat95} & 6.1 & \lsize{24}{64} & 200 & 2.56 & W--W
& & & 0.90(5) & \sl 1.32(7) \\
JLQCD & \cite{jlqcd:lat95} & 6.3 & \lsize{32}{80} & 100 & 3.38 & W--W
& & & 0.84(6) & \sl 1.24(9) \\
BBS & \cite{bbs:lat96,tblum} & 6.3 & \lsize{24}{61} & $~60$ & 3.40 &
W--W & 2 & 1.09(16) & \sl 1.01(15) & \sl 1.49(22) \\
ELC & \cite{elc:npb92} & 6.4 & \lsize{24}{60} & $~20$ & 3.7 & W--W & 3.7 &
0.86(5) & \sl 0.84(5) & \sl 1.24(7) \\
BDHS & \cite{bdhs:88} & \rlap{5.7--6.1} & & & $a{\to}0$ & W--W & 2 &
1.10(15) & \sl 0.94(14) & \sl 1.38(21) \\
BS & \cite{soni:lat95} & \rlap{5.7--6.3} & & & $a{\to}0$ & W--W & 2 &
0.96(6) & \sl 0.89(6) & \sl 1.31(8)\\[1ex]
\hline
\end{tabular}
\end{center}
\end{table*}

Recent results for the $B$ meson mixing parameter, $B_B(\mu) =
\langle\bar B| \mathcal{O}(\mu) \ket B / (8/3) f_B^2 M_B^2$, where
$\mathcal{O} =  b\gamma_\mu(1{-}\gamma_5)q \bar b
\gamma^\mu(1{-}\gamma_5)q$ is the $\Delta B=2$ operator and $M_B$ is
the $B$ meson mass come from UKQCD~\cite{ukqcd:staticB},
APE~\cite{ape:lat96} and the Kentucky group~\cite{ken:lat96} in the
static limit, together with results~\cite{soni:lat95,bbs:lat96} using
conventional methods. The APE result makes use of a new
calculation~\cite{cfg:nlo} of the full-theory/static-theory matching
which incorporates previously omitted contributions.
Table~\ref{tab:B_B} is a compilation of lattice determinations of
$B_B$. To ease comparison, values are given for $B_B(m_b{=}5\gev)$ and
for the (1-loop) renormalisation group invariant quantity, $\hat B_B =
\alpha_\mathrm{s}(\mu)^{-2/\beta_0} B_B(\mu)$. A similar collection of
results appears in the contribution by Christensen, Draper and
McNeile~\cite{ken:lat96} to these proceedings.

The conventional results are consistent and show no $a$
dependence. The static results show substantial differences arising
from the renormalisation constants. The $B_B$ calculation involves
dividing by the square of $Z_A$, the heavy-light axial current
renormalisation, which differs for Wilson (KEN) and SW (UKQCD, APE)
light quarks. Moreover, whether products of renormalisation constants
are expanded to a given order in $\alpha$ or simply multiplied makes a
significant difference: the UKQCD result for $\hat B_B$ rises to
$1.19(6)$~\cite{ukqcd:staticB} when the factors are multiplied rather
than expanded. Nonperturbative determinations of the renormalisation
factors are clearly crucial to reduce systematic
errors~\cite{gmst,marti}.

The relevant phenomenological quantity is $f_B^2 B_B$ which can be
extracted directly from the $\Delta B=2$ matrix element. To avoid
uncertainties from setting the scale, it is convenient to determine
the ratio $f_{B_s}^2 B_{B_s}/f_B^2 B_B$. $B_{B_s}/B_B$ is found to be
close to unity in recent calculations: $1.01(3)(3)$~\cite{soni:lat95}
and $1.011(8)$~\cite{ape:lat96}. For $f_{B_s}/f_B$, the values
in Table~\ref{tab:pscalar-decay-consts} can be combined with static
results,
\[
{f_{B_s}^\mathrm{stat}\over\fbstat} =
 \cases{1.11(2)&\cite{bls}\cr
1.13\erparen43&\cite{ukqcd:fBfD}\cr
1.22(4)(2)&\cite{fnal:staticB}\cr
1.16\erparen43&\cite{ukqcd:staticB}\cr
1.17(3)&\cite{ape:lat96,gimenez}\cr}
\]
The second error in the first result for $B_{B_s}/B_B$ is for
quenching based on numerical evidence for a small increase in the
ratio on $n_f=2$ dynamical configurations (Sharpe and
Zhang~\cite{sharpe-zhang,sharpe:lat96} estimate a quenching error of
$-0.04$ based on chiral loops). Unquenching is expected to increase
the value of $f_{B_s}/f_B$ by about 10\% (the chiral loop estimate in
0.16~\cite{sharpe-zhang,sharpe:lat96}).

Bernard, Blum and Soni~\cite{bbs:lat96,soni:rsd} report a preliminary
value for
\[
r_{sd} = {m_{B_s}^2 f_{B_s}^2 B_{B_s}\over m_B^2 f_B^2 B_B}
 = 1.81(8)(25).
\]
The ratio shows no evidence of $a$ dependence on a range of lattice
spacings. The result implies rather a large value for $f_{B_s}/f_B$,
compared to other lattice results: it will be interesting to see the
value of this ratio when the decay constants are extracted separately
from the same data.

\section{OTHER RESULTS}

Heavy-to-heavy quark transitions have been omitted from this
report. See~\cite{how:dubna} for a recent compilation of lattice
results for the Isgur-Wise function relevant for $B\to D^{(*)}$
semileptonic decays. UKQCD presented preliminary results at this
conference for the baryonic Isgur-Wise function in semileptonic
$\Lambda_b\to\Lambda_c$ decays~\cite{ukqcd:baryon-semilept-lat96}.

Results for $\lambda_2$, the matrix element of the chromomagnetic
moment operator between heavy mesons in the heavy quark effective
theory, continue to sit at about half the experimental
value~\cite{ape:lat96,ukqcd:staticB}. The one loop renormalisation
constant is uncomfortably large: this and other systematic effects
warrant further study given the importance of $\lambda_2$ in inclusive
$B$ decays.

Semileptonic $D\to K^{(*)}$ decays were reviewed by J.~Simone at
Lattice~95~\cite{simone:lat95}.

\subsection*{Acknowledgement}

I am very grateful to the following for supplying information about
their results: Arifa Ali Khan, Claude Bernard, Tom Blum, Joseph
Christensen, Sara Collins, Christine Davies, Giulia de Divitiis,
Vicent Gim\'enez, Shoji Hashimoto, Laurent Lellouch, Vittorio Lubicz,
Craig McNeile, Jim Simone and Nicoletta Stella. I have also received
valuable help from Patricia Ball, Aida El Khadra, Andreas Kronfeld,
Guido Martinelli, Juan Nieves, Chris Sachrajda, Amarjit Soni and
Hartmut Wittig. Where I have received new or updated results since the
St.\ Louis conference, I have endeavoured to include them in this
report, but some quoted results may already be out of date.

\bibliographystyle{elsevier}
\bibliography{lat96,ichep96}

\end{document}